\documentstyle[aps]{revtex}
\newcommand{\ds}{\displaystyle}
\newcommand{\dsf}{\ds\frac}

\newcommand{\beq}{\begin{equation}}
\newcommand{\eeq}{\end{equation}}

\large

\begin{document}
\large

\begin{center}
\Large\bf
Nonlinear Running Waves in Type II Superconductors
\vskip 0.1cm
{\normalsize\bf N.A.\,Taylanov}\\
\vskip 0.1cm
{\large\em Theoretical Physics Department and
Institute of Applied Physics,\\
National University of Uzbekistan,\\
E-mail: taylanov@iaph.post.tps.uz}
\end{center}
\begin{center}
{\bf Abstract}
\end{center}

\begin{center}
\mbox{\parbox{14cm}{\small
         A nonlinear dynamics of the thermal and electromagnetic
instabilities of critical state in type II superconductors has been
analysed taking into account the effect of dissipation and dispersion.
An existance of a nonlinear running waves describing a final stage of
evolution of the thermomagnetic instability in superconductors
is demonstrated analitically.
}}
\end{center}
\vskip 0.5cm

         The evolutions of the of the thermal $(T)$ and electromagnetic
$(\vec E, \vec H)$ perturbations is described by the nonlinear equation
of the thermal conductivity [1]
\beq
\nu\frac{dT}{dt}=\nabla [\kappa\nabla T]+\vec j\vec E,
\eeq
by the Maxwel's equations
\beq
rot\vec E=-\dsf{1}{c}\dsf{d\vec H}{dt},
\eeq
\beq
rot{\vec H}=\dsf{4\pi}{c}\vec j
\eeq
and by the equation of the resistive state
\beq
\vec j=\vec j_{c}(T,\vec H)+\vec j_{r}(\vec E).
\eeq
where $\nu=\nu(T)$ and $\kappa=\kappa(T)$ are the heat capacity and the
thermal conductivity respectively; $\vec j_c$ is the critical current density
and $\vec j_r$ is the resistance current density.

        The above system is essentially nonlinear because the right-hand
part of Eq.(1) contains a term describing the Joule heat evolution
in the region of a resistive phase.
Such a set (1)-(4) of nonlinear parabolic differential equations in
partial derivatives has no exact analytical solution [1].

        Let us consider a planar semi-infinite sample $(x>0)$ placed in an
external magnetic field $\vec H=(0, 0, H_{e})$ growing at a constant rate
$\dsf{d\vec H}{dt}=const$. According to the Maxwel's equation (2), there is a vortex
electric field $\vec E=(0, E_e, 0)$ in the sample, directed parallel to the
current density $\vec j$: $\vec E\parallel \vec j$; where $H_e$ is the
amplitude of the external magnetic field and $E_e$ is the amplitude of the
external electric field.

        A set system decision (1)-(4) in the form of functions from new
automodel variable $\xi(x,t)$ is presented:
\beq
\begin{array}{l}
T=T[\xi(x,t)],\\
\quad\\
E=E[\xi(x,t)],\\
\quad\\
j=j[\xi(x,t)].\\
\end{array}
\eeq

        Substituting (5) in a set system (1)-(4) gives as a result simple
differentation the following system
\beq
\dsf{d\xi}{dt}\left[\nu\dsf{dT}{d\xi}\right]=
\kappa\left\{\dsf{d^2\xi}{dx^2}\dsf{dT}{d\xi}+
\left(\dsf{d\xi}{dx}\right)^2
\left[\dsf{d^2T}{d\xi^2}+\left(\dsf{dT}{d\xi}
\right)^2\right]\right\}+[j_c(T)+j_n(E)]E\,,
\eeq
\beq
\dsf{d^2\xi}{dx^2}\dsf{dE}{d\xi}+\left(\dsf{d\xi}{dx}\right)^2
\dsf{d^2E}{d\xi^2}=\dsf{4\pi}{c^2}
\left[\dsf{dj_c}{dT}\dsf{dT}{d\xi}+\dsf{dj_n}{dE}\dsf{dE}{d\xi}\right]
\dsf{d\xi}{dt}\,.
\eeq

        That a set system (6),(7) at the substation (5) was only function from
$\xi$ is required performing the following conditions:
\beq
\dsf{d\xi}{dt}=A(\xi)\,,
\eeq
\beq
\left(\dsf{d\xi}{dx}\right)^2=B(\xi)\dsf{d\xi}{dt}=G(\xi)\,,
\eeq
\beq
\dsf{d^2\xi}{dx^2}=C(\xi)\dsf{d\xi}{dt}\,,
\eeq
where $A,C,G$ are the functions from $\xi$, type which will be determined
below. Solving first two system equations (8)-(10) we have a relation
\beq
G(\xi)\dsf{dA}{d\xi}=A(\xi)\dsf{dG}{d\xi}\,.
\eeq

Whence just fllows relationship between $G$ and $A$ type
\beq
G(\xi)=\dsf{1}{u}A(\xi)\,,
\eeq
where $u$  is the free constant of integrating an equation (11). From
(8) and (9) follows that $\xi(x,t)$ must satisfy single-line equation
in private derived
\beq
\dsf{d\xi}{dt}=u\dsf{d\xi}{dx}
\eeq
single deciding which is function
\beq
\xi(x,t)=F(x-ut)\,.
\eeq

        Using (14) immediately get,

\beq
G(\xi)=1,\quad A(\xi)=-Fu,\quad C(\xi)=0.
\eeq

By the transformation the coordinates and time possible to ensure
$F=1$. Thereby, definitively find sought automodel substitution
\beq
\xi=x-ut,
\eeq

corresponding to solution a type of running wave [2].

        For the automodelling solution of the form (16), describing
a running wave moving at a constant velocity $v$ along the $x$ axis,
the system of equations (1)-(4) takes the following form
\beq
- v\left[N(T)-N(T_0)\right]=\kappa\dsf{dT}{d\xi}-\frac{c^2}{4\pi v}E^2,
\eeq
\beq
\dsf{dE}{d\xi}=-\dsf{4\pi v}{c^2}j,
\eeq
\beq
E=\dsf{v}{c}H.
\eeq

       The thermal and electrodynamic boundary conditinos for equations
(17)-(19) are as follows:
\beq
\begin{array}{l}
T(\xi\rightarrow+\infty)=T_0, \dsf{dT}{d\xi}(\xi\rightarrow-\infty)=0,\\
\quad\\
E(\xi\rightarrow+\infty)=0,   E(\xi\rightarrow-\infty)=E_e,\\
\quad\\
H(\xi\rightarrow+\infty)=0,   H(\xi\rightarrow-\infty)=H_e,\\
\end{array}
\eeq
where $T_0$ is the temperature of the cooling medium.

          Let us consider the Bean-London model of the critical state for
the dependence $j_{c}(T,H)$ [3]
\beq
j_{c}(T)=j_0[1-a(T-T_{0})]
\eeq
where $j_{0}$ is the equilibtium current density,
$a$ is the thermal haet softening coefficient of the magnetic flux pinning
force.

        A characteristic field dependence of $j_r(E)$ in the region of
sufficiently strong electric fields $(E>E_f)$ can be aproximated by a
piecewise linear function $j_r\approx\sigma_f E$, where
$\sigma_f=\dsf{\eta c^2}{H\Phi_0}\approx \sigma_n H_{c_2}/H$ is the
effictive conductivity in the flux flow regime; $\eta$ is the viscous
coefficient,$\Phi_0=\dsf{\pi h c}{2e}$ is the magnetic
flux quantum, $\sigma_n$ is the conductivity in the normal state;
$E_f$ is the boundary of the linear area in the voltage-current
characteristics of sample.
In the flux creep regime $(E<E_f)$ the relation between the current density
$j$, and the electric field, $E$, is strongly nonlinear [4].

     Excluding are variables $T(\xi)$ and $H(\xi)$ using Eqs.(17) and (19) ,
and taking into account the boundary conditions (20), we obtain an equation
describing the electric field $E(\xi)$ distribution ($E$-wave):
\beq
\dsf{d^2 E}{d\xi^2}+\left[\dsf{4\pi v}{c^2}\dsf{dj_r}{dE}\dsf{dE}{d\xi}+
\dsf{4\pi v^2a}{c^2}\dsf{N(T)-N(T_0)}{\kappa (T)}\right]-
\dsf{aE^2}{2\kappa (T)}=0,
\eeq
where the dependency $T=T\left(E,\dsf{dE}{d\xi}\right)$ is defined by
expression (2), (4) and have the form

\beq
T=T\left(E,\dsf{dE}{d\xi}\right)=T_0+\dsf{1}{a}\left[j_0+j_r(E)+
\dsf{4\pi v}{c^2}\dsf{dE}{d\xi}\right].
\eeq

Here $N(T)=\int\limits_{0}^{T}\nu(T)dT$.

The analysis of the phase plane ($E,\dsf{dE}{dz}$) of Eq.(22) shows that
there are two equilibrium points: $E_0=0$, $T=T_0$ is the stabile node and
$E=E_e$, $T=T*=T(E_e,0)$ is the saddle. The sepatetrix, joining these
two equilibrium points represents a solution of the shock-wave-type with
amplitude $E_e$ (see, [7]).
The velocity of an $E$-wave is determined by the Eq. (22) with account
a boundary conditions (20):
\beq
v_{E}^{2}=\dsf{c^2}{8\pi}\dsf{E_{e}^{2}}
{N\left[T_0+\dsf{1}{a}[j_c(T)+j_r(E)]\right]-N(T_0)}.
\eeq
Using Eqs. (19) and (22) we find the expression for a distribution of the
magnetic field $H$ in the case of $H$-wave :
\beq
\dsf{d^2H}{d\xi^2}+\dsf{4\pi v}{c^2}\left[\left.\dsf{dj_r}{dE}
\right|_{E=\dsf{v}{c}H}\dsf{dH}{d\xi}+ca\dsf{N(T)-N(T_0)}
{\kappa (T)}\right]-\dsf{av}{2c}\dsf{H^2}
{\kappa (T)}=0.
\eeq

Скорость $H$ - волны $v_H$ связана с ее амплитудой следующим соотношением
\beq
N(T)-N(T_0)=\dsf{H_{e}^{2}}{8\pi}.
\eeq
The condition (9) reflects tha adiabatical character of the wave propagation:
the magnetic field energy transferred by the wave ensures local heating of
the sample in the close vicinity of the wave front.
Thereby the depending on the external conditions at the sample surfase
a possible to exist two kinds thermomagnetic waves.

Note that within the linear voltage-current chatacteristics $(E>E_f)$
in the approximation of weak superconductor heating $(T-T_0)<<T_c-T_0)$,
the heat capacity $\nu$ and the thermal conductivity $\kappa$
depend on temperature profile only slightly. Then in the limiting case
$\tau>>1$ [5] the effect of dispersion on the wave dynamics is
negigible. Correspondingly equation (22) coincides with the first
integral of Burgers equation [6], which have solution of the form [7]
\beq
E(z)=\dsf{E^*}{2}\left[1-th\dsf{\beta}{2} (z-z_0)\right].
\eeq

Here the following dimensionless parameters are introduced
$E^*=2\beta^2\tau E_\kappa$,
$\beta=\dsf{vt_\kappa}{L}$ and $z=\dsf{\xi}{L}$; where
$L=\dsf{cH_e}{4\pi j_0}$ is the depth of the magnetic field penetration
into the superconductor,
$t_\kappa=\dsf{\nu L^2}{\kappa}$ is the time of thermal diffusion,
$\tau=\dsf{D_t}{D_m}$ is the parameter describing
the ratio of the thermal $D_{t}=\dsf{\kappa}{\nu}$ to the magnetic
$D_{m}=\dsf{c^2}{4\pi\sigma_{f}}$ diffusion coefficient;
$E_{\kappa}=\dsf{\kappa}{aL^2}$ is a constant.

        Using the boundary condition $E(z\rightarrow-\infty)=E_e$
we can be determined the velocity $v_{E}$
\beq
v_E=\dsf{L}{t_\kappa}\left[\dsf{E_e}{2\tau E_\kappa}\right]^{1/2},
\eeq
the front width
\beq
\delta z=16\dsf{(1+\tau)}{\tau^{1/2}}\left[\dsf{E_\kappa}{E_e}\right]^{1/2}.
\eeq
and the time of a wave propagating inside the sample
\beq
\Delta t=\dsf{L}{2v_E}.
\eeq
Numerical estimations give $v_E\approx 1\div 10^2 \dsf{sm}{sec}$,
$\delta z\approx 10^{-1}\div 10^{-3}$ and $\Delta t\approx 10^{-4}$ sec
for characteristics values of the physical parameters [5].

The system Eqs. (1)-(4) is invariant with respect to an arbitrary translation.
Therefore, the wave propagation conditions can be found for an arbitrary
critical current density dependence on $T$ and $H$. The results can also be
obtained for an arbitrary temperature dependence of thermophysical parameters
$\nu$ and $\kappa$ of superconducting material and for an arbitrary function
$j_r(E)$.

The problem of stability of nonlinear shock-wave with
respect to small thermal and electromagnetic perturbations is studied in [7].
It is also shown that only damped perturbations correspond to space-limited
solutions, which means that nonlinear wave is stable.

  I would like to thank I. Maksimov for stimulating discussions.

\newpage
\centerline{\large \bf NIZAM A.TAYLANOV}
\begin{tabbing}
{\bf Address:} \\
Theoretical Physics Department of National University of Uzbekistan,\\
Vuzgorodok, 700174, Tashkent, Uzbekistan\\
Telephone:(9-98712),461-573, 460-867.\\
fax: (9-98712) 463-262,(9-98712) 461-540,(9-9871) 144-77-28\\
e-mail: taylanov@iaph.post.tps.uz \\
\end{tabbing}
\end{document}